\documentstyle[aps,preprint]{revtex}

\draft

\begin{document}

\title{{\bf Thermodynamic Properties of a Trapped Interacting Bose Gas}}
\author{{\it Hualin Shi and Wei-Mou Zheng} \\
{\it Institute of Theoretical Physics, Academia Sinica, Beijing 100080, China%
}}
\date{31Oct, 1996}
\maketitle

\begin{abstract}
A Bose gas in an external potential is studied by means of the local
density approximation. Analytical results are derived for the 
thermodynamic properties of an ideal Bose gas in a generic 
power-law trapping potential, and their dependence on the mutual
interaction of atoms in the case of a non-ideal Bose gas.
\end{abstract}


One of the most striking consequences of quantum statistics is the 
condensation of an ideal Bose-Einstein gas where the zero momentum state can 
become macroscopically occupied at a sufficiently low 
temperature\cite{huang87}. Some of the features of the transition to a 
superfluid phase exhibited by the Bose fluid $^4$He is often interpreted 
essentially as a result of this Bose-Einstein condensation (BEC), the 
strongest degeneracy effect of a Boson system\cite{pen56}. For
many years, it was considered hopeless to experimentally observe BEC in an
atomic gas with weak interactions. With the development of techniques to
trap and cool atoms, BEC was recently observed directly in dilute atomic
vapors\cite{and95,bra95,dav95}. The new experimental achievements have
stimulated great interest in the theoretical study of inhomogeneous Bose
gases.

The thermodynamic properties of trapped atomic Bose gases undergoing BEC can
be altered by the spatially varying trapping potential. The interaction
between atoms may have a significant effect on the thermodynamic properties.
There have been several investigations analyzing the dependence of the
critical temperature on the trapping potential and weak interaction in the
Bose gas\cite{bag87,gio96,shi96b}. The thermodynamic properties of Bose 
gases in an external potential have also been discussed in Ref.~\cite{bag87}, 
but no analytic results have been given. Here we shall derive some analytical 
expressions for the thermodynamic properties of a trapped non-interacting and 
interacting Bose gas under the local density approximation.

Let us first consider the case of an ideal Bose gas trapped in an external
potential. When the energy level spacing due to the trapping potential is much
smaller than the thermal energy $k T = \beta^{-1}$, the local density 
approximation is adequate\cite{gol81,oli89,chou96}. Space may then be divided 
into small cells, and in each cell we may consider the trapping potential 
$V({\bf r})$ to be constant. With a trivial modification to indicate $r$ 
dependence explicitly we may directly extend the formula of the grand 
potential, which can be found in any textbook on statistical mechanics, 
e.g. Ref.~\cite{huang87}, to write the local grand potential above the 
critical temperature $T_c$ as 
\begin{equation}
\label{p1-den}\widetilde{\Omega }_{BE}({\bf r})=-\frac{kT}{\lambda ^3}%
g_{5/2}(\zeta )
\end{equation}
where $\mu$ is the chemical potential, the thermal wave length $\lambda = 
\left( {2\pi \hbar ^2}/{m k T}\right)^{\frac 1 2}$, and 
\begin{equation}
\label{def}\zeta =\exp [-\beta (V({\bf r})-\mu )], \qquad g_\nu
(x)=\sum_{j=1}^\infty j^{-\nu }x^j.
\end{equation}
From this grand potential the local density of the Bose gas 
in the external potential is
\begin{equation}
\label{n-den}
\rho (T,{\bf r})=-\left(\frac{\partial \widetilde{\Omega }_{BE}({\bf r})}
{\partial \mu }\right)_T=\frac 1{\lambda ^3}g_{3/2}(\zeta ) .
\end{equation}
This is the same as \cite{chou96}. For the generic power-law potential 
discussed in Ref.~\cite{bag87}
\begin{equation}
\label{pot}V({\bf r})=\epsilon _1\left| \frac x{L_1}\right| ^p+\epsilon
_2\left| \frac y{L_2}\right| ^l+\epsilon _3\left| \frac z{L_3}\right| ^q,
\end{equation}
where $L_1$, $L_2$ and $L_3$ are the linear sizes of the volume, and 
$\varepsilon_1$, $\varepsilon_2$, $\varepsilon_3$, $p$, $l$ and $q$ the 
parameters, integrating the local grand potential (\ref{p1-den}) over the 
whole volume, we find the total grand potential of 
the system to be
\begin{eqnarray}
\Omega _{BE}(T,\mu ) &=&-kT\lambda ^{-3}\int g_{5/2}(\zeta )
d{\bf r} \nonumber\\  
&=&-{\strut \frac W{\lambda^3\beta ^{\eta +1/2}}}\; g_{\eta +2}(z),
\label{pt1-den}
\end{eqnarray}
where $z=e^{\beta\mu}$, $\eta =1/p+1/l+1/q+1/2$, and  
\begin{equation}
W=\frac{8L_1L_2L_3\;I(p,l,q)}{\varepsilon _1^{1/p}
\varepsilon_2^{1/l}\varepsilon _3^{1/q}},\quad I(p,l,q)=(plq)^{-1}\;\Gamma (1/p)
\;\Gamma (1/l)\;\Gamma (1/q).
\end{equation}
In the derivation we have used the definition of the function $g_\nu (x)$ 
and the formula for the gamma function
\begin{equation}
\label{gamma}
\Gamma (z)=\int_0^\infty t^{z-1}e^{-t}\;dt.
\end{equation}
From the total grand thermodynamic potential the average atom number can be 
obtained as
\begin{eqnarray}
\langle N\rangle  &=&-\left( {\strut \frac{\partial \Omega _{BE}}
{\partial \mu }}\right)_T \nonumber\\
  &=&{\strut\frac W{\lambda^3\beta ^{\eta -1/2}}}\; z{\strut
  \frac{\partial }{\partial z} }g_{\eta +2}(z)\nonumber\\
  &=&{\strut\frac W{\lambda ^3\beta ^{\eta -1/2}}}g_{\eta +1}(z) .
\label{Nav}
\end{eqnarray}
This is in agreement with the result in Ref.~\cite{shi96b} obtained 
directly from the local density $\rho (r)$ instead of the grand partition 
function. 

The total entropy can be obtained as
\begin{eqnarray}
\langle S\rangle  &=& -\left(
{\strut \frac{\partial \Omega _{BE}}{\partial T}}\right)_\mu\nonumber  \\  
&=&{\strut\frac {Wk}{\lambda^3\beta ^{\eta -1/2}}}
\left[(\eta +2)g_{\eta +2}(z)-\beta\mu g_{\eta +1}(z)\right].
\label{st1}
\end{eqnarray}
Although, for obvious reasons, the volume and pressure are not useful 
thermodynamic variables for a trapped gas\cite{bag87}, we may still consider 
an analogue to the capacity for the homogeneous system at constant volume 
with a fixed number of atoms, and define this heat capacity as
\begin{equation}
C(T)=T\left( \frac{\partial S}{\partial T}\right)_N,
\label{c1-def}
\end{equation}
which includes the work done against the potential as the energy of the gas
is changed. 

So far we have discussed the case for a temperature above $T_c$. When the 
temperature $T$ is below the BEC critical temperature $T_c$, the
chemical potential $\mu =0$, and only the normal component of the Bose gas 
has a contribution to the heat capacity. From Eqs.~(\ref{st1}) and 
(\ref{c1-def}) we can then obtain the heat capacity ($T<T_c$) to be:
\begin{eqnarray}
C_- (T) & =&{\strut\frac{(\eta +1)(\eta +2)Wk}{\lambda ^3\beta 
^{\eta -1/2}}}\;g_{\eta +2}(1) \nonumber\\
& =&{\strut\frac{(\eta +1)(\eta +2)\;g_{\eta +2}(1)}{g_{\eta +1}(1)}}
{\strut \left( \frac{T}{T_c} \right) }^{\eta +1}\langle N\rangle k.
\label{cm1}
\end{eqnarray}
As a limit, at the critical temperature $T_c$, the above expression reduces 
to
\begin{equation}
C(T_c^-) =\frac{(\eta +1)(\eta +2)\;g_{\eta +2}(1)}{g_{\eta +1}(1)}
\langle N\rangle k.
\label{cm1p}
\end{equation}

When the temperature $T$ is above the critical temperature $T_c$, at a 
fixed $N$ the chemical potential $\mu $ is a function of temperature $T$. 
In order to obtain the heat capacity, we have to calculate the derivative 
of $\mu $ with respect to temperature $T$ while holding $N$ fixed. Using
\begin{equation}
\frac{\partial \langle N\rangle }{\partial T}=0,
\end{equation}
from Eq.~(\ref{Nav}) we find
\begin{equation}
(\eta +1) g_{\eta +1}(z)+\left[\frac 1 k 
\left(\frac{\partial \mu}{\partial T}\right)
_{\langle N\rangle }-\beta\mu \right]g_\eta (z)=0,
\end{equation}
which yields 
\begin{equation}
\left( \frac{\partial \mu }{\partial T}\right) _{\langle N\rangle }=\frac
\mu T-\frac{k(\eta +1)g_{\eta +1}(z)}{g_\eta (z)}.
\label{dm1}
\end{equation}
From Eqs.~(\ref{c1-def}) and (\ref{dm1}) the heat capacity above $T_c$ is 
then given by
\begin{eqnarray}
C_+(T) & =&{\strut\frac {Wk}{\lambda ^3\beta ^{\eta -1/2}}}\left[ (\eta +1)
(\eta +2)g_{\eta +2}(z)-(\eta +1)^2 g_{\eta +1}^2(z)/g_\eta 
(z)\right]\nonumber  \\  
& =&\left[ (\eta +1)(\eta +2){\strut\frac{g_{\eta +2}(z)}{g_{\eta +1}(z)}}
-(\eta +1)^2\;{\strut\frac{g_{\eta +1}(z)}{g_\eta (z)}}\right] 
\langle N\rangle k.
\label{ct1}
\end{eqnarray}
At $T=T_c^{+}$, the chemical potential $\mu =0$, so the above expression 
reduces to 
\begin{equation}
C(T_c^{+})=\left[ (\eta +1)(\eta +2)\frac{g_{\eta +2}(1)}{g_{\eta +1}(1)}%
-(\eta +1)^2\frac{g_{\eta +1}(1)}{g_\eta (1)}\right] \langle N\rangle k.
\label{cp1}
\end{equation}
By comparing with Eq.~(\ref{cm1p}), the discontinuity of heat capacity at the 
critical point is
\begin{equation}
\Delta C=C(T_c^{-})-C(T_c^{+})=\frac{(\eta +1)^2g_{\eta +1}(1)}{g_\eta (1)}
\langle N\rangle k.
\label{dc1}
\end{equation}
From the Eqs.~(\ref{cm1}), (\ref{cp1}) and (\ref{dc1}), we see that the heat 
capacity depends only on the particle number and the exponents $p$, $l$ and 
$q$ of the external potential. We can further examine the continuity of $C(T)$ 
and its derivatives. As we know, $g_{\eta}(1)$ is divergent for $\eta \leq 1$.
So, the heat capacity becomes discontinuous at the critical temperature $T_c$ 
when $\eta >1$. From Eqs.~(\ref{cm1}) and (\ref{ct1}), the derivative 
$\partial C(T) /\partial T$ is discontinuous at $T_c$ for either $\eta \leq 1$
 or $\eta >1$. This is in agreement with Ref.~\cite{bag87}. From the 
analytical expressions we also see that the heat capacity at the critical point 
grows with increasing $\eta$.
 
For a rigid wall box, $\eta=1/2$. It is straightforward to check that the 
above expressions indeed reduce to the known results in textbooks of 
statistical mechanics (e.g. Ref.~\cite{huang87}). As a special case let us 
consider the harmonic potential
\begin{equation}
V({\bf r})=\frac{1}{2} m\omega _{\perp }^2r_{\perp }^2+\frac{1}{2}
m\omega _z^2r_z^2 .
\label{harmonic}
\end{equation}
We find the heat capacity $C(T_c^+)$, $C(T_c^-)$ and $\Delta C$ to be
\begin{eqnarray}
C(T_c^{-})&=&{\strut\frac{12 g_{4}(1)}{g_{3}(1)}}\langle N\rangle k,\\
C(T_c^{+})&=&\left[ {\strut\frac{12 g_{4}(1)}{g_{3}(1)}} -{\strut\frac{
9 g_3(1)}{g_2(1)}} \right] \langle N\rangle k,\\
\Delta C&=& {\strut\frac{9 g_3(1)}{g_2(1)}} \langle N\rangle k.
\end{eqnarray}
This agrees with Refs.~\cite{gro50,bag87}.

Along similar lines we may analyze the thermodynamic properties of a 
trapped interacting Bose gas. In the mean field approximation, the single 
particle effective potential can be written as:
\begin{equation}
V_{eff}({\bf r})=V({\bf r})+2 a \lambda^2 \rho ({\bf r}),
\end{equation}
where $a$ is the $s$-wave scattering length, which characterizes the strength
of interaction between two atoms. We assume $|a|\ll\lambda$ for the weak 
interaction considered here. By similarity with Eq.~(\ref{p1-den}), 
for the trapped interacting Bose gas the local grand potential can be written 
as
\begin{equation}
\label{p2-den}\widetilde{\Omega }_{BE}^I({\bf r})=-kT\lambda^{-3}\; 
g_{5/2}(\xi ) ,
\end{equation}
where
\begin{equation}
\xi =\exp [-\beta (V({\bf r})+2a\lambda ^2\rho ({\bf r})-\mu )] .
\end{equation}
Expanding the function $g_{5/2}(\xi )$ in Eq.~(\ref{p2-den}) according to 
the small parameter $a/\lambda$, and keeping terms only up to the lowest 
order in $a/ \lambda$, we have
\begin{eqnarray}
\widetilde{\Omega }_{BE}^I({\bf r}) & \simeq& -
kT\lambda ^{-3}\left[ g_{5/2}(\zeta )-2a\lambda ^2\rho ({\bf r}
)g_{3/2}(\zeta )\right]\nonumber \\  
& \simeq& -kT\lambda ^{-3}\left[
g_{5/2}(\zeta )-2a\lambda ^{-1}g_{3/2}^2(\zeta )\right],
\end{eqnarray}
which coincides with the first order result of the virial expansion of the grand 
partition
function\cite{huang57}. This is not surprising since the mean field 
approximation is accurate to the first order. From the grand potential we 
can obtain the local density
\begin{eqnarray}
\rho^I ( {\bf r}) & =&-\left( {\strut\frac{\partial}{\partial \mu }} \widetilde
{\Omega }_{BE}^I ({\bf r}) \right)_T \nonumber\\  
& =&\lambda ^{-3}\left[ g_{3/2}(\zeta )-4a\lambda^{-1}g_{1/2}(\zeta )
g_{3/2}(\zeta )\right] \nonumber\\
& \simeq& \lambda ^{-3}g_{3/2}(\exp [-\beta (V({\bf r})+4a\lambda ^2\rho (
{\bf r})-\mu )]),
\end{eqnarray}
which, compared with the density of an ideal Bose gas, means the effective 
fugacity is $\exp[-\beta (V({\bf r})+4a\lambda ^2\rho ({\bf r})-\mu )]$. 
However, from a naive consideration based on the effective interaction one 
would think that it should be $\exp[-\beta (V({\bf r})+2a\lambda ^2\rho 
({\bf r})-\mu )]$ instead. The factor two in the atomic density expression 
naturally appears by taking the derivative of the grand potential with respect 
to $\mu$.

As the counterpart to Eq.~(\ref{pt1-den}), we can find the total grand 
potential for the interacting gas at $T>T_c$ to be
\begin{equation}
\Omega _{BE}^I=-\frac W{\lambda ^3\beta ^{\eta +1/2}}\left[
g_{\eta +2}(z)-2a\lambda ^{-1}F_{3/2,\;3/2,\;\eta }(z)\right],
\label{pt2-den}
\end{equation}
where
\begin{equation}
F_{\delta ,\nu ,\eta }(x)=\sum_{i,j=1}^\infty \frac{x^{i+j}}{i^\delta j^\nu
(i+j)^{\eta -1/2}}.
\end{equation}
When $a=0$, Eq.~(\ref{pt2-den}) recovers expression (\ref{pt1-den}) for the 
non-interacting Bose gas, as it should. The average number of atoms is 
now given by
\begin{eqnarray}
\label{nt2}
\langle N^I\rangle  & =&-\left({\strut \frac{\partial \Omega _{BE}^I}
{\partial \mu }}\right)_T \nonumber\\  
& =&{\strut \frac W{\lambda ^3\beta ^{\eta -1/2}}}\left[ g_{\eta +1}(z)-
2a\lambda^{-1}F_{3/2,3/2,\eta -1}(z)\right],
\end{eqnarray}
where we have used
\begin{equation}
\frac{\partial F_{\delta ,\nu ,\eta }(x)}{\partial x}=\frac{1}{x}\;
F_{\delta ,\nu,\eta -1}(x).
\end{equation} 
As we know,  $g_{\eta+1}(z)$ is a bounded, positive, monotonically 
increasing function of $z$, and $F_{\delta ,\nu ,\eta }(x)$ is positive.
From Eq.~(\ref{nt2}) we see that the trapped interacting Bose gas has a 
higher chemical potential than an ideal Bose gas when the atomic number and 
temperature are the same in both systems.

From Eq.~(\ref{nt2}) for $\langle N^I\rangle$ we can calculate the derivative 
$\partial \mu/\partial T$ at a fixed $N^I$, and the result is
\begin{eqnarray}
\label{dmu2}
\left( {\strut\frac{\partial \mu }{\partial T}}\right)_{\langle N^I\rangle } 
& =&{\strut\frac\mu T}- {\strut\frac{(\eta +1)g_{\eta +1}(z)-2a
\lambda ^{-1}(\eta +3/2)F_{3/2,3/2,\eta -1}(z)}{g_\eta (z)-2a\lambda ^{-1}
F_{3/2,3/2,\eta -2}(z)}} k \nonumber\\  
& \simeq&{\strut\frac \mu T} -\left\{ {\strut\frac{(\eta +1)g_{\eta +1}(z)}
{g_\eta (z)}}+{\strut\frac{2a}\lambda} \right[ {\strut\frac{(\eta +1)
g_{\eta +1}(z)F_{3/2,\;3/2,\;\eta -2}(z)}{g_\eta^2(z)}}\nonumber\\
&&-\left.\left.{\strut
\frac{(\eta +3/2)F_{3/2,\;3/2,\;\eta -1}(z)}{g_\eta (z)}}\right] \right\}k.
\end{eqnarray}
Corresponding to Eq.~(\ref{st1}), the total entropy of the trapped 
interacting Bose gas can then be derived as
\begin{eqnarray}
\label{entropy2}
\langle S^I\rangle  & =&-\left({\strut\frac{\partial \Omega _{BE}^I}
{\partial T}}\right)_\mu \nonumber \\  
& =&{\strut\frac {Wk}{\lambda ^3\beta ^{\eta -1/2}}}\left[(\eta +2)g_{\eta 
+2}(z)
-\beta\mu \;g_{\eta +1}(z)\right]\nonumber \\
& &\quad -{\strut\frac{2aWk}{\lambda ^4\beta ^{\eta -1/2}}}\left[ (\eta +5/2)
F_{3/2,\;3/2,\;\eta }(z)-\beta\mu\;F_{3/2,\;3/2,\;\eta -1}(z)\right].
\end{eqnarray}
Calculating the derivative of the entropy $\langle S^I\rangle$, we obtain 
the heat capacity above the BEC critical temperature to be 
\begin{eqnarray}
\label{heat2}
C^I (T) & =&T\left({\strut\frac{\partial S^I}{\partial T}}
\right)_{N^I} \nonumber\\  
& =&\left[ (\eta +1)(\eta +2){\strut\frac{g_{\eta +2}(z)}{g_{\eta +1}(z)}}
-(\eta +1)^2{\strut\frac{g_{\eta +1}(z)}{g_\eta (z)}}\right] 
\langle N\rangle k -{\strut
\frac{2a}\lambda }\left[ {\strut\frac{(\eta +3/2)(\eta +5/2) 
F_{3/2,\;3/2,\;\eta}(z)}{g_{\eta+1} (z)}} \right. \nonumber\\
&& \quad\left. +(\eta +1)\left( {\strut\frac{(\eta +1)g_{\eta+1} (z)
F_{3/2,\;3/2,\;\eta -2}(z)}{g_\eta ^2(z)}}-{\strut\frac{2(\eta +3/2)
F_{3/2,\;3/2,\;\eta -1}(z)}{g_{\eta}(z)}}\right)\right] \langle N\rangle k.
\end{eqnarray}
The last two formulas show the correction due to the mutual interaction of 
atoms to thermodynamic quantities such as the entropy and heat capacity. 

The above derivations for the trapped interacting Bose gas is adequate only 
for the gas phase (without BEC), and furnish no information whatsoever 
about the condensed phase, or the nature of the 
condensation\cite{huang57}. Any attempts to use the same equations in 
the regime of BEC will lead to some mistakes. According to 
Ref.~\cite{huang57} the mean field effective potential should be amended
to include the contributions of atoms to Bose condensation. If
we choose an appropriate effective mean field potential similar
to Eq.~(28) of Ref.~\cite{huang57}, we can, in principle, extend the 
above discussion to calculate thermodynamic quantities below the 
critical temperature. However, the derivation will be so involved that 
no simple analytical formulas can be expected to exist any more.

\acknowledgments
{The authors thank Hao Bai-lin for his encouragement and useful
discussions. This work was supported in part by the National
Natural Science Foundation of China.}

\end{document}